\documentstyle[12pt]{article}
\newcommand{\ii}{\'{\i}}

\newcommand{\prodin}{\mbox{$\displaystyle \prod_{i=1}^{N}$}}

\newcommand{\comeq}{\begin{equation}}
\newcommand{\comeqvec}{\begin{eqnarray}}
\newcommand{\tereq}{\end{equation}}
\newcommand{\tereqvec}{\end{eqnarray}}

\newcommand{\intinf}{\mbox{$\displaystyle \int_{-\infty}^{+\infty}$}}
\newcommand{\etai}{\mbox{$\eta_{i}$}}
\newcommand{\chii}{\mbox{$\chi_{i}$}}
\newcommand{\etais}{\mbox{$\eta_{i}^{*}$}}
\newcommand{\chiis}{\mbox{$\chi_{i}^{*}$}}
\newcommand{\etaj}{\mbox{$\eta_{j}$}}
\newcommand{\chij}{\mbox{$\chi_{j}$}}

\newcommand{\chijs}{\mbox{$\chi_{j}^{*}$}}

\newcommand{\wcuad}{\mbox{$\mid\omega\mid^{2}$}}
\newcommand{\bb}{\bibitem}
\newcommand{\nn}{\nonumber}

\begin{document}

\begin{titlepage}
\title{Distribution of eigenvalues of ensembles of
asymmetrically diluted Hopfield matrices}
\vspace{2cm}
\author{D. A. Stariolo,
	 E. M. F. Curado\thanks{Present address:
                      Universidade Federal de Bras\ii lia,
		      Brazil}
        \, and F. A. Tamarit\thanks{Present address:
                      FaMAF, Universidad Nacional de C\'ordoba,
                      Argentina}  \\
Centro Brasileiro de Pesquisas F\ii sicas/CNPq \\
 Rua Xavier Sigaud 150\\
  22290-180 Rio de Janeiro--RJ, Brazil}
 \date{}
\maketitle

\vspace{2cm}

\noindent {\bf Abstract. }
Using Grassmann variables and an analogy with two dimensional
electrostatics, we obtain the average eigenvalue distribution
$\rho(\omega)$ of ensembles of $N \times N$ asymmetrically
diluted Hopfield matrices in the limit
$N \rightarrow \infty$. We found that in the limit of strong
dilution the distribution is uniform in a circle in the complex
plane.
\end{titlepage}
%

Random matrix theory has become an active field of research in
mathematics and physics in the last decades. In physics, since the now
classical work of Metha\cite{me:67} on the statistical
description of the energy levels of atomic nuclei, random
matrices have emerged  as an important tool in the study of the
localization transition\cite{fy:mi:91,ef:83}, quantum chaos\cite{ha:91},
spin glasses\cite{bi:yo:86}, neural networks\cite{am:89,solla:91}, and
disordered systems in general. Most of the work deal with ensembles of
hermitian or symmetric matrices whose individual properties are
well known  and can be exploited in more complex situations. In
the last years  the interest in asymmetric matrices has grown,
motivated, besides its intrinsic mathematical value,
 by problems of dissipative quantum
dynamics\cite{gr:ha:so:88} and the dynamics of neural
networks\cite{so:cr:so:88}. In a recent paper, Sommers
et.al.\cite{so:cr:so:st:88} calculated the average density of
eigenvalues $\rho(\omega)$
of $N \times N$  random asymmetric matrices in the limit
$N \rightarrow \infty$, with  elements $J_{ij}$,
given by a Gaussian distribution with zero mean and correlations
\begin{equation}
N\langle\langle J_{ij}^{2}\rangle\rangle_{J} \; =1, \hspace{1cm}
N\langle\langle J_{ij} \; J_{ji}\rangle\rangle_{J} \; = \; \tau
\end{equation}
They found that the eigenvalues are uniformly distributed inside an
ellipse in the complex plane, whose  semi axes
depend on the degree of
asymmetry of the ensemble $\tau$. Generalizing this result,
Lehmann {\em et. al.} \cite{le:so:91} calculated the joint probability
distribution of eigenvalues in Gaussian ensembles of real
asymmetric matrices, recovering the elliptic law in the large N limit.

In this paper we calculate the average eigenvalue distribution
$\rho(\lambda)$ of an ensemble of asymmetrically diluted Hopfield matrices,
whose elements are given by:
\begin{equation}
J_{ij}=\frac{c_{ij}}{N} \sum_{\mu=1}^{p} \xi_{i}^{\mu} \xi_{j}^{\mu}
\hspace{2cm} i,j=1\ldots N \; ,
\end{equation}
where $\{\xi_{i}^{\mu}\, i=1\ldots N, \mu=1\ldots p\}$ represents
a set of p
random patterns. The $\xi_{i}^{\mu}$ are random independent
variables that can take the values $\pm 1$  with the same probability
and the $c_{ij}$ are random variables chosen accordingly
to the following distribution
\begin{equation}
P(c_{ij})=\gamma\delta(c_{ij}-1)+(1-\gamma)\delta(c_{ij}).\label{cij}
\end{equation}
$0\leq\gamma\leq 1$ measures the degree of dilution of the
matrices. $\gamma=1$ corresponds to symmetric Hopfield
matrices whose eigenvalue distribution is known\cite{cr:so:87}.

In order to obtain the distribution $\rho(\omega)$ we use an
analogy  with a two dimensional electrostatic problem introduced
in ref.\cite{so:cr:so:st:88}. Let us define the Green function
associated to the matrix $J$
\begin{equation}
G(\omega)=\frac{1}{N}\; {\Large \langle\langle} Tr\frac{1}{I\omega-J}
{\Large \rangle\rangle}_{J} \; ,
\end{equation}
where $\omega=x+iy$ is a complex variable, $I$ the identity
matrix and $\langle \langle \dots \rangle \rangle_{J}$  denotes an
average over the random variables $\xi_{i}^{\mu}$. If $\lambda_{i},\,
i=1,\ldots ,N$ are
the eigenvalues of $J$, then
\begin{equation}
Tr\frac{1}{I\omega-J}=\sum_{i}^{N} \frac{1}{\omega-\lambda_{i}}
\end{equation}
For large $N$ the sum can be approximated by an integral and the
Green function becomes
\begin{equation}
G(\omega)=
\int d^{2}\lambda \frac{\rho(\lambda)}{\omega-\lambda}
\end{equation}
where $\rho(\lambda)$ is the density of eigenvalues in the plane.
The last equation suggests an analogy with a two-dimensional
classical electrostatics problem in which $\rho(\lambda)$ represents the
density of charge in the plane. It can be
demonstrated\cite{so:cr:so:st:88} that an electrostatic
potential $\Phi$ exists, satisfying
\begin{equation}
2 Re G = -\frac{\partial\Phi}{\partial x}, \hspace{2cm}
-2Im G = -\frac{\partial\Phi}{\partial y}
\end{equation}
and which obeys Poisson's equation:
\begin{equation}
\nabla^{2}\Phi=-4\pi\rho
\end{equation}
Thus, in order to determine $\rho(\omega)$
we may calculate the potential $\Phi$.  Using that
$det(AB)=detA detB$ and $detA^{T}=detA$ one can prove that a good
definition for $\Phi$ can be
\begin{equation}
\Phi(\omega)=-1/N{\Large\langle\langle} \ln \det\left( (I\omega^{*}-J^{T})
             (I\omega -J)\right){\Large\rangle\rangle}_{J}
\end{equation}
with $\omega^{*}$ the complex conjugate of $\omega$ and $J^{T}$
the transpose of $J$.
In what follows we will consider the case $N \rightarrow \infty$
and assume that in this limit the average and the $\ln$
operations commute \cite{so:cr:so:st:88}.
By using a Grassmannian representation\cite{grass} of the
determinant and adding a matrix $\epsilon \; \delta_{ij}$, with
$\epsilon$ positive and infinitesimal in order to avoid zero
eigenvalues, we get:
\begin{eqnarray}
\lefteqn{ \exp\left[-N\Phi(\omega)\right]=
 {\Large\langle\langle} \int_{-\infty}^{\infty}
 \left( \prod_{i=1}^{N} d\eta_{i}^{*} \; d\eta_{i} \right) } \nn \\
&&   \exp\left\{ -\sum_{i,j,k}\eta_{i}^{*}
(\omega^{*}\delta_{ik}-J_{ik}^{T})
(\omega \delta_{kj}-J_{kj} ) \eta_{j}
-\epsilon \sum_{i} \eta_{i}^{*} \eta_{i} \right\}
{\Large\rangle\rangle}_{J}
\end{eqnarray}
After performing the average over the $c_{ij}$ and
over the random patterns $\{\xi_{i}^{\mu}\}$, we arrive at the
following expression:
\begin{eqnarray}
\lefteqn{ \exp\left[-N\Phi(\omega)\right]=}  \nonumber \\
&& \int_{-\infty}^{\infty} \left( \prod_{i=1}^{N} d\eta_{i} d\eta_{i}^{*}
   \right)
   \exp \left\{ (\epsilon+\mid \omega \mid^{2})Nq  -
   \alpha N \ln{t}
   + \alpha \gamma(\omega+\omega^{*}) Nq-
   \alpha \gamma(1-\gamma)\wcuad Nq^{2} \right\} \nn  \\
&& \times \intinf \left( \prodin d\chiis d\chii \right) \exp \left\{
   -\sum_{i} \chiis\chii [1+\alpha\gamma(1-\gamma)q] + \right. \nn \\
&& \left. \sum_{i}\chiis\etai [\alpha\gamma-\alpha\gamma(1-\gamma)\omega q] +
   \sum_{i} \etais\chii [\alpha\gamma -
   \alpha\gamma(1-\gamma)\omega^{*} q] \right\} \nn \\
&& \times \exp \alpha N \ln \left\{ 1-\gamma^{2}qt\sum_{i} \chiis\chii /N
   - \gamma t (\sum_{i} \chiis\etai+\sum_{i} \etais\chii)/N + \right. \nn \\
&& \left. \gamma^{2} t (\sum_{i} \etais\chiis\sum_{j}\etaj\chij +
   \sum_{i}\etais\chii \sum_{j} \chijs\etaj)/N^{2} \right\} \label{expNfi}
\end{eqnarray}
where
\begin{eqnarray}
q& =& \frac{1}{N}\sum_{i}\eta_{i}^{*}\eta_{i} \\
t& =& \frac{1}{1-\frac{\gamma}{N}(\omega+\omega^{*}) \sum_{i}
\eta_{i}^{*}\eta_{i}}
\end{eqnarray}
respectively.  Next we define the
following parameters:
\begin{eqnarray}
q=\frac{1}{N}\sum_{i=1}^{N} \eta_{i}^{*} \eta_{i}  & &\hspace{2cm}
z=\frac{1}{N}\sum_{i=1}^{N} \chi_{i}^{*} \chi_{i} \nn \\
r=\frac{1}{N}\sum_{i=1}^{N} \eta_{i}^{*} \chi_{i}  & &\hspace{2cm}
r^{*}=\frac{1}{N}\sum_{i=1}^{N} \chi_{i}^{*} \eta_{i} \\
s=\frac{1}{N}\sum_{i=1}^{N} \eta_{i}^{*} \chi_{i}^{*} & &\hspace{2cm}
s^{*}=\frac{1}{N}\sum_{i=1}^{N}\eta_{i}\chi_{i} \nn
\end{eqnarray}
and  introduce them into Eq. (\ref{expNfi}) by using
delta functions. After integrating over the Grassmann
variables we get:

\begin{eqnarray}
\lefteqn{\exp \left[ -N\Phi(\omega) \right]= } \nn \\
        && \left( \frac{N}{2\pi} \right)^{4}
        \int_{-\infty}^{\infty} dq\,dQ\,dz\,dZ\,dr\,dR\,
        dr^{*}\,dR^{*} \nn \\
        && \times \exp N \large \left\{
        qQ+zZ+rR+r^{*}R^{*}+
	[\epsilon+\mid \omega \mid^{2} +
        \alpha\gamma(\omega+\omega^{*})] q \right.\nn \\
	&& +\alpha\ln [1-\gamma(\omega+\omega^{*})q]
        +  \ln(RR^{*}-ZQ)-z+
           \alpha\gamma(r+r^{*}) \nn \\
        && -\alpha\gamma(1-\gamma)[\wcuad q^{2}+zq+\omega
           r^{*}q+\omega^{*}rq] \nn \\
        && \left.+\alpha\ln\left[1-\frac{\gamma^{2}qz}
           {1-\gamma(\omega+\omega^{*})q}-
           \frac{\gamma(r+r^{*})}{1-\gamma(\omega+\omega^{*})q}+
\frac{\gamma^{2}rr^{*}}{1-\gamma(\omega+\omega^{*})q}\right]\right\}
\end{eqnarray}
In the large $N$ limit this multiple integral can be evaluated
by the saddle point method.
Up to now the calculation is exact for arbitrary
$\gamma$. Since the resulting  saddle point equations are
difficult to solve analytically, in this work we present the
results for the strong dilution limit
$(\gamma \ll 1)$. Expanding the exponent in powers of $\gamma$
and keeping terms up to $O(\gamma)$ we obtain, after some calculations:
\begin{eqnarray}
\lefteqn {\exp \left[ -N\Phi(\omega) \right] \propto} \nn \\
&& \intinf dq \exp -N \left[ \ln \mid q \mid - (\epsilon+\wcuad)q +
\alpha\gamma \mid \omega \mid^{2} q^{2} - \ln (1+\alpha\gamma q) \right]
\end{eqnarray}
After a change of variables $\sigma=1/q$ we arrive at the
following saddle-point equation:
\begin{equation}
\frac{\epsilon}{\sigma^{2}}=\frac{1}{\sigma+\alpha\gamma}-
\frac{x^{2}}{(\sigma+\alpha\gamma)^{2}}-
\frac{y^{2}}{(\sigma+\alpha\gamma)^{2}} \label{saddle}
\end{equation}
Expanding $\epsilon$ in powers of $\sigma$,
 the solution of the
saddle-point equation in the limit $\epsilon\rightarrow 0$
is $\sigma=0$ inside the circle
$x^{2}+y^{2}=\alpha\gamma$. In this region
$G(\omega)=\omega^{*}/(\alpha\gamma)$ (non analytic) and
$\nabla^{2}\Phi=-4/(\alpha\gamma)$.
This implies that the density of eigenvalues is uniform inside a
circle of radius $\sqrt{\alpha\gamma}$ in the complex plane.
Outside the circle the solution to Eq.(\ref{saddle}) becomes
$\sigma=x^{2}+y^{2}-\alpha\gamma$, the Green function is
$G(\omega)=1/\omega$ (analytic), and the density $\rho=0$. The
density of eigenvalues in the whole complex plane is:
\begin{eqnarray}
\rho(\omega)= \left\{ \begin{array}{ll}
                      1/ \pi\alpha\gamma & if\;\;\;\; x^{2}+y^{2}\le
\alpha\gamma \\
		      0                  & otherwise
		      \end{array} \right. \label{ro}
\end{eqnarray}
It is important to note that $\langle\langle J_{ij}J_{ji} \rangle \rangle_{J}
\propto
\gamma^{2}$ and consequently, in
this limit of strong dilution, the matrix elements
become effectively uncorrelated and we obtain a
``circular law''. This result has to be compared with the similar result of
ref.\cite{so:cr:so:st:88} for the case $\tau=0$, i.e., a
Gaussian ensemble of completely asymmetric random matrices.
It is expected that this circle deforms into
an ellipse as the asymmetry parameter $\gamma$ increases and
permits the appearance of random correlations between the patterns.

In figures (1-3) we show the results of
numerical diagonalization of sets of $N \times N$  matrices, with linear
sizes $N$ ranging from $512$ to $2048$ for $\alpha=0.25$
and $\gamma=0.01$.
The figures show the projection of the
distribution of complex eigenvalues in the real
axis. The full curves represent the analytical solution:
\begin{eqnarray}
\rho_{x} &=& \int \rho(x,y) dy \nonumber \\
         &=& \frac{2}{\pi\alpha\gamma} (\alpha \gamma - x^2)^{1/2},
       \hspace{2cm} \mid x \mid \leq \sqrt{\alpha\gamma}\; .
\end{eqnarray}
We found that the numerical results present a
peak at the origin that becomes smaller as the size N increases.
Assuming that it is a finite size effect, and that the weight of
the peak might be uniformly distributed on the whole support of
the distribution, we renormalized the distributions. After
renormalizing the numerical data the agreement with the analytic
curves becomes very good as the size increases. Figure (4) shows
the dependence of the peak at the origin with the system size.
We have fitted the data at the origin with an exponential
function in $1/N$, $\rho_x(0)=a\exp{(b/N)}$. The extrapolation to
$N \rightarrow \infty$ coincides with an error of $10 \%$ with the
analytic result at the origin.

Concluding, in this paper we have obtained analytically the
distribution of eingenvalues of an ensemble of asymmetrically
diluted Hopfield matrices in the limit of strong dilution.
The eingenvalues are uniformly distributed inside a circle in
the complex plane. Our results are supported by numerical
diagonalization of the ensemble considered. Although we
presented only the results for the strong dilution limit, the
saddle point equations are valid for any amount of dilution.
Work for solving the general case is in progress.

\vspace{2cm}
{\Large Acknowledgment}\\
We wish to thank Prof. Luca Peliti for valuable discussions and
the ``Centro Nacional de Supercomputa\c{c}\~{a}o da Universidade
Federal do Rio Grande do Sul'' for permitting us the use of the
Cray Y-MP2E/232 where the numerical part of the work was done.

\newpage
{\Large Caption for figures}
\begin{description}
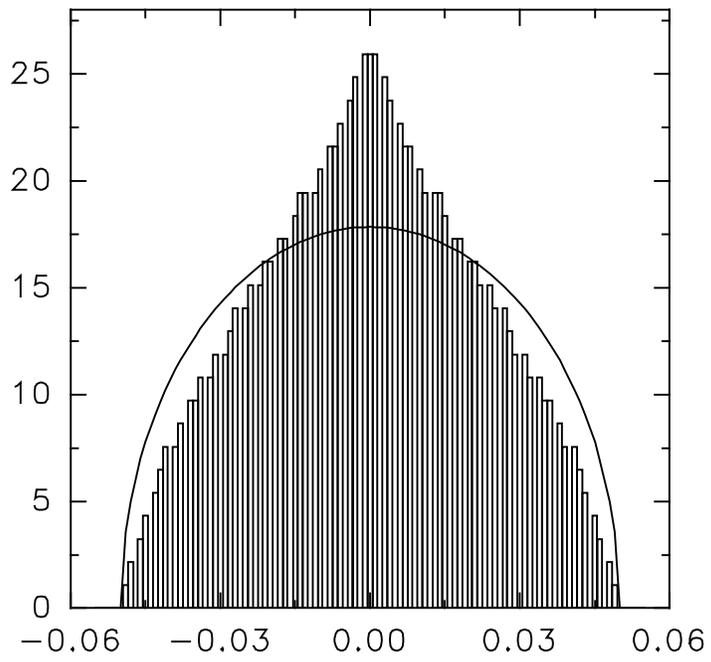
\item [Figure 1] Projection of the eingenvalue distribution on
the real axis. The
full line correspond to the analytical results and the histogram
to the numerical diagonalization performed with $\gamma=0.01$,
$\alpha=0.25$ and $N=512$ and averaged over 20 realizations of
the matrices.
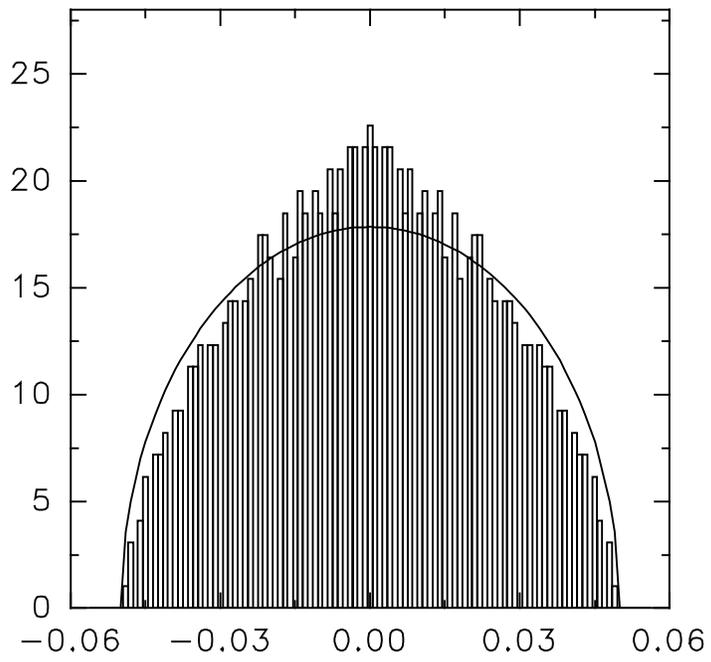
\item [Figure 2] Same as figure (1) with $N=1024$ and averaged
over 10 realizations of the matrices.
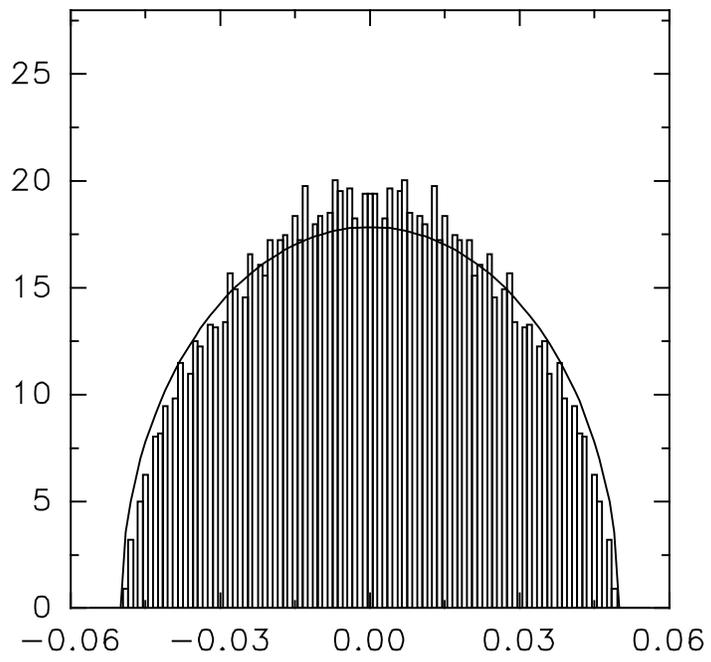
\item [Figure 3] Same as previous figures with $N=2048$ averaged
over 8 realizations of the matrices.
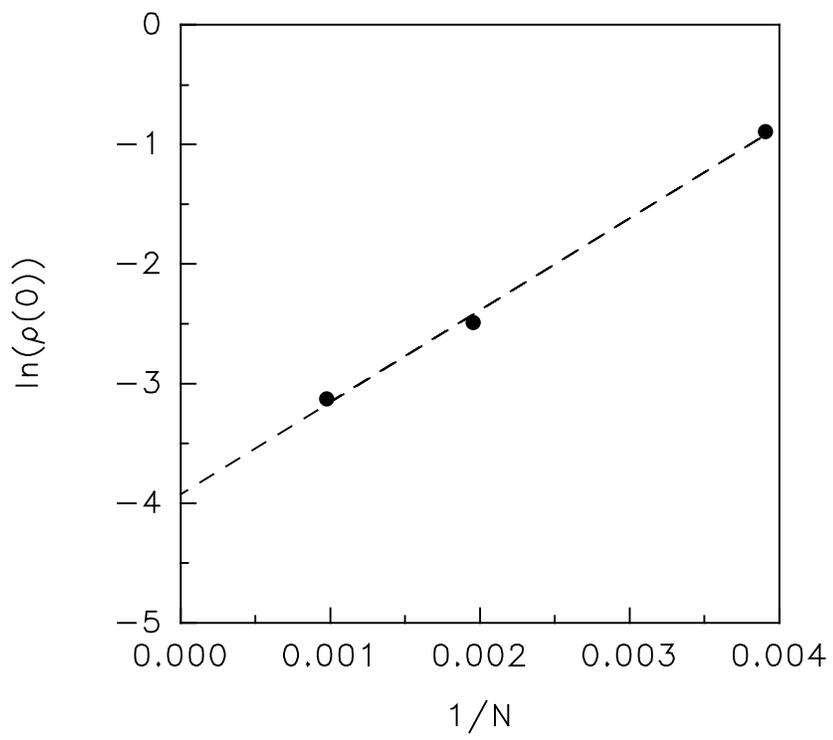
\item [Figure 4] Finite size scaling of the peak at the origin
in the complex plane $\ln{\rho(0)}$ versus $1/N$.
\end{description}


%

\end{document}